\documentclass[aps,prl,superscriptaddress,amsmath,amssymb,showpacs,twocolumn,notitlepage]{revtex4-1}
\usepackage{amssymb} \usepackage{amsmath} \usepackage{graphicx}
\usepackage{bm} \usepackage{bbm} \usepackage{color} \usepackage{times}
\usepackage{xcolor} 

\renewcommand{\eqref}[1]{(\ref{#1})}
\newcommand{\Eqref}[1]{Equation (\ref{#1})}

\graphicspath{{./Figures/}}

\newcommand{\citeasnoun}[1]{Ref.~\onlinecite{#1}}

\begin{document}

\title{Strongly coupled near-field radiative and conductive heat transfer between planar objects}

\author{Riccardo Messina}
\thanks{These authors contributed equally to this work.}
\affiliation{Laboratoire Charles Coulomb (L2C), UMR 5221 CNRS-Universit\'{e} de Montpellier, F- 34095 Montpellier, France}
\author{Weiliang Jin}
\thanks{These authors contributed equally to this work.}
\affiliation{Department of Electrical Engineering, Princeton University, Princeton, NJ 08544, USA}
\author{Alejandro W. Rodriguez}
\affiliation{Department of Electrical Engineering, Princeton University, Princeton, NJ 08544, USA}

\begin{abstract}
  We study the interplay of conductive and radiative heat transfer
  (RHT) in planar geometries and predict that temperature gradients
  induced by radiation can play a significant role on the behavior of
  RHT with respect to gap sizes, depending largely on geometric and
  material parameters and not so crucially on operating
  temperatures. Our findings exploit rigorous calculations based on a
  closed-form expression for the heat flux between two plates
  separated by vacuum gaps $d$ and subject to arbitrary temperature
  profiles, along with an approximate but accurate analytical
  treatment of coupled conduction--radiation in this geometry. We find
  that these effects can be prominent in typical materials
  (e.g. silica and sapphire) at separations of tens of nanometers, and
  can play an even larger role in metal oxides, which exhibit moderate
  conductivities and enhanced radiative properties. Broadly speaking,
  these predictions suggest that the impact of RHT on thermal
  conduction, and vice versa, could manifest itself as a limit on the
  possible magnitude of RHT at the nanoscale, which asymptotes to a
  constant (the conductive transfer rate when the gap is closed)
  instead of diverging at short separations.
\end{abstract}

\pacs{}

\maketitle

Two non-touching bodies held at different temperatures can exchange
heat through photons. In the far field, i.e. separations $d \gg$
thermal wavelength $\lambda_T=\hbar c/k_BT$ (of the order of 8\,$\mu$m
at room temperature), the maximum radiative heat transfer (RHT)
between them is limited by the well-known Stefan--Boltzmann
law~\cite{howell2010thermal}. In the near field, i.e.  $d \ll
\lambda_T$, evanescent waves can tunnel and contribute flux, enabling
RHT to exceed this limit by several orders of
magnitude~\cite{basu2009review,volokitin2007near,joulain2005review}. Such
enhancements can be larger in nano-structured
surfaces~\cite{miller2015shape,khandekar2015giant}, but only a handful
of these have been studied thus
far~\cite{narayanaswamy2008thermal,kruger2011nonequilibrium,mccauley2012modeling,rodriguez2013fluctuating},
leaving much room for improvements~\cite{miller2015shape}. A more
commonly studied heat-transport mechanism is thermal conduction,
involving transfer of energy through phonons or electrons. Although
conduction is typically more efficient than
RHT~\cite{holman2010heat,wong2011monte,lau2016parametric}, there are
ongoing theoretical and experimental efforts aimed at discovering
novel materials and structures leading to larger RHT, with recent work
suggesting the possibility of orders of magnitude
enhancements~\cite{miller2015shape}, in which case RHT could not only
compete but even exceed conduction in situations typically encountered
in everyday experiments (e.g. under ambient conditions).

In this paper, we present an approach for studying coupled
conduction--radiation (CR) problems between planar objects separated
by gaps that captures the full interplay of near-field RHT and thermal
conduction at the nanoscale. Using an exact, closed-form expression of
the slab--slab RHT in the presence of arbitrary temperature
distributions, we show that CR interplay can give rise to significant
temperature gradients and thereby greatly modify RHT, causing the
latter to asymptote to a constant, the conductive flux when the gap is
closed, which sets a fundamental limit to radiative heat exchange at
short separations. We provide evidence of the validity of a simple but
useful surface--sink approximation that treats the impact of RHT on
conduction as arising purely at the vacuum--slab interfaces, yielding
analytical expressions with which one can study the scaling behavior
and impact of these effects with respect to relevant parameters. For
instance, we find that their prominence is largely independent of
operating temperatures but strongly tied to the choice of materials
(e.g. glasses and oxides versus highly conductive metals) and
geometries (e.g. thin versus macroscopic films). Furthermore, these
phenomena lie within the reach of current-generation
experiments~\cite{ShenNanoLetters09}, leading to significant changes
in RHT between typical materials like silica and saphire at relatively
large gap sizes $\sim$ several tens of nanometers, and potentially
playing an even greater role in metal oxides, which exhibit low-loss
infrared polaritons~\cite{chiritescu2007ultralow,cahill2014nanoscale}
and therefore enhance RHT.

Coupled photonic and phononic diffusion processes in nanostructures
are becoming increasingly
important~\cite{JoulainJQSRT,chiloyan2015natcomm}. While recent works
have primarily focused on the interplay between thermal diffusion and
external optical illumination, e.g. laser-induced, localized heating
of plasmonic
structures~\cite{baffou2010mapping,baffou2014deterministic,ma2014heat,baldwin2014thermal,biswas2015sudden},
the thermal radiation emitted by a heated body and absorbed by nearby
objects can also be a great source of heating or cooling. To date,
however, the impact of RHT on conduction remains largely unexplored,
with the consensus view being that radiation is insufficiently large
to result in appreciable temperature
gradients~\cite{wong2011monte,wong2014coupling,lau2016parametric}. On
the other hand, modern experiments measuring RHT between planar
surfaces are beginning to probe the nanometer
regime~\cite{KittelPRL05,HuApplPhysLett08,NarayanaswamyPRB08,RousseauNaturePhoton09,ShenNanoLetters09,KralikRevSciInstrum11,OttensPRL11,vanZwolPRL12a,vanZwolPRL12b,KralikPRL12,KimNature15,StGelaisNatureNano16,SongNatureNano15,KloppstecharXiv},
and in certain cases offer evidence of deviations from the typical
$1/d^2$ behavior associated with near-field
enhancement~\cite{ChapuisPRB08,MuletMTE02} at nanometric distances,
often attributed to nonlocal~\cite{Henkel06,JoulainJQSRT} or
phonon-tunneling~\cite{chiloyan2015natcomm} effects. Here, we find
that depending on geometric configuration and materials,
radiation-induced temperature gradients can play a significant role on
transport above the nanometer regime, requiring a full treatment of
the coupling between conduction and radiation.

\emph{Exact formulation of coupled conduction--radiation.---} In what
follows, we present a formulation of coupled CR applicable to the
typical situation of two planar bodies (the same framework can be
extended to multiple bodies), labelled $a$ and $b$, separated by a gap
of size $d$. We assume that the slabs exhibit arbitrary temperature
profiles and exchange heat among one other. Neglecting convection and
considering bodies with lengthscales larger or of the order of their
phonon mean-free path, in which case Fourier conduction is valid, the
stationary temperature distribution satisfies the one-dimensional
coupled CR equation:
\begin{equation}
  \frac{\partial}{\partial z} \left[ \kappa(z)\frac{\partial}{\partial
      z} T(z)\right]+\int \mathrm{d}z'\,\varphi(z',z)=Q(z),
\label{eq:heat}
\end{equation}
where $\kappa(z)$ and $Q(z)$ represent the bulk Fourier conductivity
and external heat-flux rate at $z$,
respectively, while $\varphi(z',z)$ denotes the radiative power per
unit volume from a point $z'$ to $z$. Previous studies of
\eqref{eq:heat} considered only radiative energy escaping into vacuum
through the surfaces of the objects, exploiting simple, albeit
inaccurate ray-optical approximations that are inapplicable for
sub-wavelength objects or in the near
field~\cite{da2004conduction,holman2010heat}. The novelty of our
approach to \eqref{eq:heat} is that $\varphi$, as written above and
computed below, is fundamentally tied to accurate and modern
descriptions of RHT based on macroscopic fluctuational
EM~\cite{rytov1988principles,basu2009review}, allowing us to explore
regimes (e.g. distances $\ll\lambda_T$) where near-field effects
dominate RHT among different objects. In particular, as we show below,
in some regimes RHT can lead to observable temperature
distributions. Although we only consider the impact of external
radiation on the temperature profile and vice versa, under large
temperature gradients, RHT could potentially modify the intrinsic
thermal conductivity of these
objects~\cite{chen2005surface,JoulainJQSRT,volz2016nanophononics}, a
situation that we leave to future work. We also ignore far-field
radiation since it is negligible compared to conduction or RHT at the
distances considered in this work.

\begin{figure}[t!]
\begin{center}
\includegraphics[height=6.1cm]{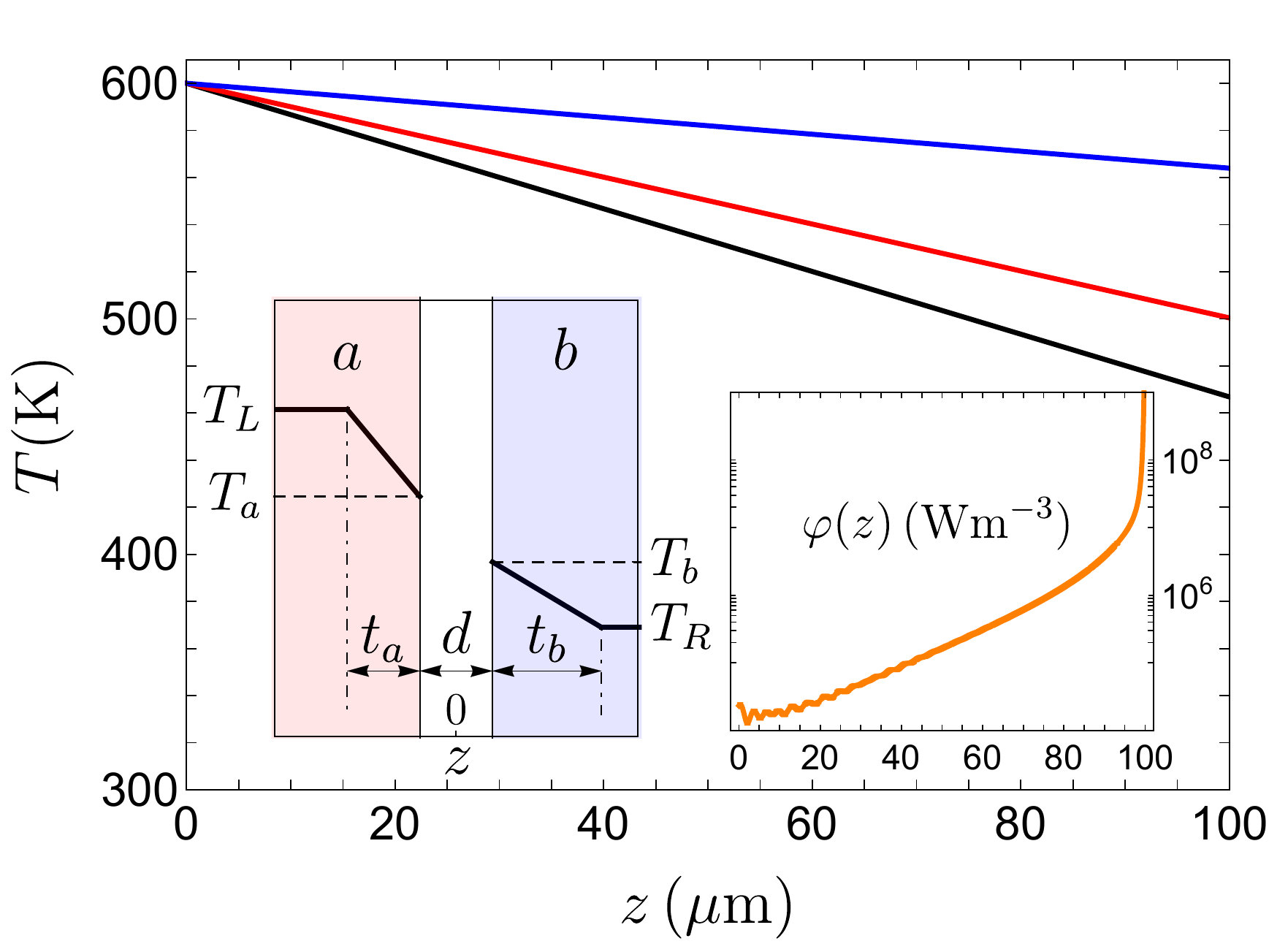}
\end{center}
\caption{The left inset shows a schematic of two parallel slabs
  separated by a distance $d$ along the $z$ direction ($z=0$ denotes
  the middle of the gap). The two slabs exhibit a temperature profile
  $T(z)$: the temperature is constant and has values $T_L$ and $T_R$
  in the regions $z\leq -d/2-t_a$ and $z \geq d/2+t_b$, respectively,
  and variable in the regions $-d/2-t_a<z<-d/2$ and $d/2<z<d/2+t_b$,
  with $T_a$ and $T_b$ denoting the temperatures at the slab--vacuum
  interfaces. In the main part of the figure, temperature profile
  along the temperature-varying region of the hot slab in a
  configuration involving two silica slabs with $t_a=t_b=100\,\mu$m
  held at external temperatures $T_L=600\,$K and $T_R=300\,$K at the
  outer ends, and separated by vacuum gaps $d$. The points
  $z=0,100\,\mu$m represent the boundary of the thermostat and the
  interface with the vacuum gap. The three lines correspond to
  $d=10\,$nm (black), 20\,nm (red) and 50\,nm (blue). The right inset
  shows the $z$-dependent radiative flux-rate $\varphi(z_a)$ (see
  text) along slab $a$ at $d=100\,$nm.}
\label{validity}
\end{figure}

We focus on the scenario illustrated on the inset of
Fig.~\ref{validity}, in which the temperature of slab $a$ ($b$) slab
is fixed at $T_L$ ($T_R$) by means of a thermostat, except for a
region of thickness $t_a$ ($t_b$). To study this problem, we calculate
the RHT via a Fourier expansion of the slabs' scattering
matrices~\cite{francoeur2009solution}. Such techniques were recently
employed to obtain close-formed analytical expressions of RHT between
plates of uniform
temperature~\cite{ben2009near,MessinaPRA11,Messina2012prl,MessinaPRA14}. Here,
we extend these results to consider the more general problem of slabs
under arbitrary temperature distributions. Toward this aim, we
generalize prior methods~\cite{MessinaPRA11,MessinaPRA14} by dividing
each slab into films of infinitesimal thicknesses, each having a fixed
temperature. Describing the associated EM fields at each point by
means of the fluctuation-dissipation theorem (from which the RHT can
be deduced), and considering multiple reflections associated with the
various interfaces, we find that the evanescent RHT per unit volume
from a point $z_a$ in slab $a$ to a point $z_b$ in slab $b$,
$\varphi(z_a,z_b)=\int_0^{\infty}\mathrm{d}\omega\int_{\omega/c}^{\infty}\mathrm{d}\beta~\varphi_a(\omega,\beta;z_a,z_b)$,
can be expressed analytically in the closed form~\cite{futurework}:
\begin{equation}\label{eq:slab}\begin{split}
 \varphi(\omega,\beta;z_a,z_b)&=\frac{4\beta}{\pi^2}(r''k_{zm}'')^2\frac{e^{-2k_z''d}\,e^{-2k_{zm}''(z_b-d/2)}}{|1-r^2e^{-2k_z''d}|^2}\\
 &\,\times\Bigl(N[\omega,T(z_a)]-N[\omega,T(z_b)]\Bigr),
\end{split}
\end{equation}
where $\beta$ denotes the conserved, parallel ($x$--$y$) wavevector
$k_z=\sqrt{\omega^2/c^2-\beta^2}$ and
$k_{zm}=\sqrt{\varepsilon\omega^2/c^2-\beta^2}$ the perpendicular
wavevectors in vacuum and the interior of the slabs, respectively, and
where we introduced the transverse-magnetic (dominant) polarization
Fresnel reflection coefficient of a planar $\varepsilon$--vacuum
interface, $r=(\varepsilon k_z-k_{zm})/(\varepsilon k_z+k_{zm})$. Note
that we restrict our analysis to the transverse-magnetic polarization
because only it supports a surface phonon-polariton
resonance. \Eqref{eq:slab} permits fast solutions of coupled CR
problems in this geometry for a wide range of parameters.

\emph{Surface--sink approximation:} At small separations, near-field
RHT is dominated by large-$\beta$ surface modes that are exponentially
confined to the slab--vacuum
interfaces~\cite{joulain2005review}. Hence, it is sufficient (as
discussed below and confirmed through exact results in
Fig.~\ref{validity}) to treat its impact on conduction as a purely
surface effect, in which case the entire problem can be described
through the surface temperatures. In particular, under this
assumption, given a distance $d$, identical conductivities $\kappa$
and temperature-varying regions $t=t_a=t_b$, and external temperatures
$T_L$ and $T_R$, the only unknowns are the interface temperatures
$T_a$ and $T_b$, which satisfy the following boundary conditions:
\begin{equation}\label{bound}
 -\kappa\frac{T_a-T_L}{t}=-\kappa\frac{T_R-T_b}{t}=\varphi,
\end{equation}
Here, $\varphi$ denotes the net heat exchanged between the two slabs
(assumed to take place at the surfaces), whose spectral $\omega$ and
$\beta$ components are given by~\cite{futurework}:
\begin{equation}\begin{split}
 &\varphi(\omega,\beta)=\frac{2\beta}{\pi^2}(r'')^2k_{zm}''\frac{e^{-2k_z''d}}{|1-r^2e^{-2k_z''d}|^2}\int_0^{+\infty}\!\!\!\!dz\,e^{-2k_{zm}''z}\\
 &\times\Bigl(N[\omega,T(-d/2-z)]-N[\omega,T(d/2+z)]\Bigr).
\end{split}
\end{equation}
Despite the complex dependence of the heat flux on separation and
temperature profile, we find that it is possible to approximate the
former using a simple, power-law expression of the form, $\varphi
\simeq h_0(T_a-T_b)/d^2$~\cite{MuletMTE02} (valid as long as the
radiation is primarily coming from the surface of the slabs), with the
coefficient $h_0$ calculated as the near-field heat flux between two
uniform-temperature slabs held at $T_L$ and $T_R$, divided by
$T_L-T_R$. Essentially, while the dependence of RHT on absolute
temperature is generally nonlinear, the fact that conduction through
the interior of the slabs scales linearly with $T_a-T_b$ and that
energy must be conserved (i.e. changes in conductive transfer must be
offset by corresponding changes in RHT), implies that $\varphi$ must
also scale linearly with $T_a-T_b$, with the precise value of the
coefficient $h_0$ determined from the radiative conductivity at large
values of $d$ where radiation does not impact conduction. Given these
simplifications, Eq.~\eqref{bound} can be solved to yield:
\begin{equation}
\label{analytics}
\frac{T_a-T_b}{T_L-T_R}=\left(1+\frac{2th_0}{\kappa
  d^2}\right)^{-1}\!\!\!,\,\frac{\varphi}{T_L-T_R}=\frac{h_0}{d^2} \left(\frac{T_a-T_b}{T_L-T_R}\right).
\end{equation}
These formulas reveal that the interplay of conduction and radiation
causes $T_a-T_b \to 0$ quadratically with $d$, producing a continuous
temperature profile and leading to a finite value of $\varphi \to
\kappa(T_L-T_R)/2t$ as $d \to 0$, the conductive flux through a
gapless slab of thickness $2t$ subject to a linear temperature
gradient $T_L-T_R$ (as it must, from energy conservation). Below, we
show that the existence of such temperature gradients along with
deviations from the typical $1/d^2$ RHT power law are within the reach
of present experimental detection.

\emph{Numerical predictions.---} To begin with, we first address the
validity of the surface--sink approximation above. In order to do so,
we of course need to consider the full coupled CR problem described
by~\eqref{eq:heat}, requiring numerical evaluation of the spatial heat
transfer $\varphi(z_a,z_b)$ in~\eqref{eq:slab}. For concreteness, we
consider a practical situation typical of RHT experiments, involving
two silica (SiO$_2$) slabs subject to external temperatures
$T_L=600\,$K and $T_R=300\,$K by a thermostat at distance
$t=t_a=t_b=100\,\mu$m away from the slab--vacuum interfaces. Silica
not only has relatively low $\kappa\approx 1.4\,$W/m$\cdot$K but also
supports polaritonic resonances at mid-infrared wavelengths and has
well-tabulated optical properties~\cite{Palik98}.
Figure~\ref{validity} illustrates the increasing, \emph{linear}
temperature gradient present in slab $a$ with decreasing separations
$d$, a consequence of the exponential decay of the spatial heat
transfer, $\varphi(z_a) = \int dz_b\, \varphi(z_b,z_a)$, illustrated
on the inset at a fixed $d=100\,$nm. Results obtained
through~\eqref{analytics}, with $h_0=5.53\times10^{-12}\,$W/K, are in
almost perfect (essentially indistiguishable) agreement with those of
the full CR treatment and are therefore not shown. The same is true at
smaller values of $t$, down to tens of nanometers, below which the
surface--sink approximation begins to fail.

\begin{figure}[t!]
\begin{center}
\includegraphics[height=6.1cm]{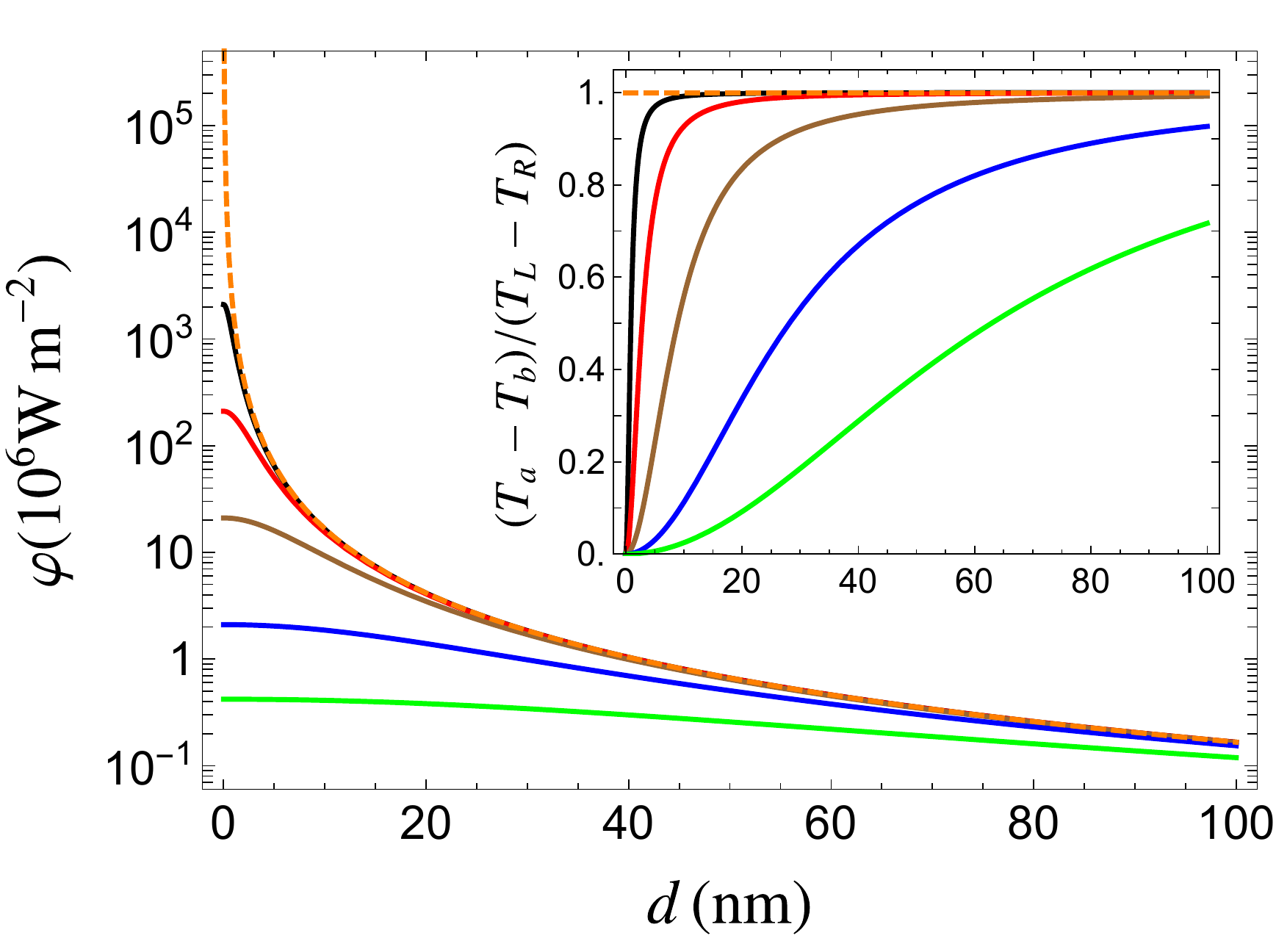}
\end{center}
\caption{Total flux $\varphi$ and temperature difference $T_a-T_b$
  (inset) as a function of distance $d$ between two silica slabs
  (shown schematically on the left inset of Fig.~\ref{validity}) that
  are being held at $T_L=600\,$K and $T_R=300\,$K. The various solid
  lines correspond to different temperature-varying regions $t$ (from
  top to bottom): 100\,nm (black), $1\,\mu$m (red), $10\,\mu$m
  (brown), $100\,\mu$m (blue) and $500\,\mu$m (green). The orange
  dashed line shows $\varphi$ in the absence of temperature
  gradients.}
\label{silica}
\end{figure}

Figure~\ref{silica} shows $\varphi$ and $T_a-T_b$ (inset), normalized
by the external temperature difference $T_L-T_R$, as a function of $d$
and for the same slab configuration but considering multiple
$t=\{0,0.1,1,10,100,500\}\,\mu$m, with decreasing values of $t$
leading to smaller temperature gradients and larger $\varphi$. Here,
$t=0$ (dashed line) corresponds to the typical scenario where
conduction dominates and hence there are no temperature gradients, in
which case $\varphi = h_0(T_L-T_R)/d^2$ exhibits the expected
divergence. Quite interestingly, we find that at typical values of
$t=100\,\mu$m, the flux decreases by $\approx 50\%$ at distances
$d\approx 30\,$nm, well within the reach of current
experiments~\cite{KittelPRL05,ShenNanoLetters09,KloppstecharXiv,SongNatureNano15,KimNature15}. This
result may be particularly relevant to recent
experiments~\cite{ShenNanoLetters09} investigating RHT between large
silica objects, which indicate deviations from the $1/d^2$ scaling
behavior (along with flux saturation) at similar distances.

\begin{figure}[t]
\begin{center}
\includegraphics[height=6.1cm]{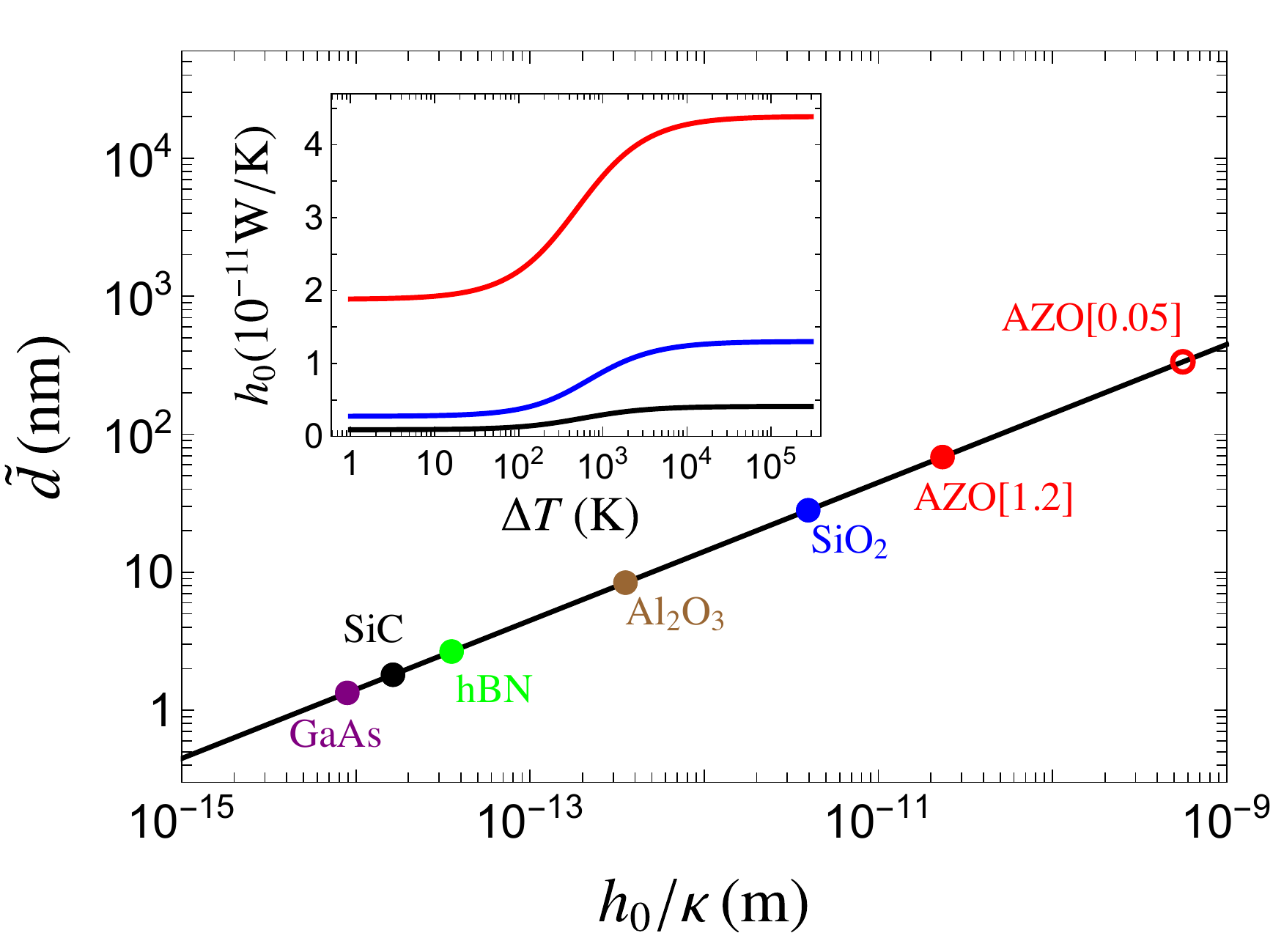}
\end{center}
\caption{Typical distance scale $\tilde{d}$ relevant to
  conduction--radiation problems (see text) as a function of the
  material-dependent ratio $h_0/\kappa$ for two slabs with
  $t_a=t_b=100\,\mu$m. Solid circles denote corresponding values for
  specific material choices (abbreviations), with AZO[1.2] and
  AZO[0.05] denoting aluminum zinc oxides of different conductivities
  $\kappa=1.2\,$W/m$\cdot$K~\cite{loureiro2014transparent} and
  potential $\kappa=0.05\,$W/m$\cdot$K~\cite{chiritescu2007ultralow},
  respectively. The inset shows the dependence of the radiative-heat
  transfer coefficient $h_0$ on the external temperature gradient
  $\Delta T=T_L-300\,$K, for three cases, AZO (red), silica (blue) and
  SiC (black), with decreasing values.}
\label{h0}
\end{figure}

We now explore the degree to which these saturation effects depend on
the choice of material and operating conditions, quantified via the
separation regime at which they become significant. In particular,
inspection of \eqref{analytics} alllows us to define the distance
$\tilde{d} = \sqrt{2th_0/\kappa}$ at which
$T_a-T_b=\frac{1}{2}(T_L-T_R)$ and $\varphi = \frac{1}{2}
h_0(T_L-T_R)/\tilde{d}^2$, corresponding to half the value of the RHT
obtained when conduction and radiation do not influence one
another. Figure~\ref{h0} shows $\tilde{d}$ as a function of the
material-dependent ratio $h_0/\kappa$ for the particular choice of
$T_L=600$~K, $T_R=300$~K, and $t=100\,\mu$m, highlighting the
square-root dependence of the former on the latter. Superimposed are
the expected $\tilde{d}$ associated with various materials of possible
experimental interest (solid circles), obtained by employing
appropiate values of $\kappa$ and $h_0$, which depend primarily on the
choice of external temperature. Within the surface--sink approximation
(valid here), the latter do not influence the scaling of either
$\varphi$ or $T_a-T_b$ with respect to separation, as evident from
\eqref{analytics}. The inset of Fig.~\ref{h0} shows $h_0$ as a
function $T_L$ for SiC, SiO$_2$, and aluminum zinc oxide (AZO),
identified by their increasing values of $h_0$, illustrating the near
constancy of the coefficient over a wide range of acceptable
temperature differences. Note that we consider unrealistically large
values of $T_L$ only to illustrate asymptotic behavior.

Noticeably, despite small differences in the value of $h_0$ between
various materials, there are striking variations in $\tilde{d}$, which
can range anywhere from a few nanometers in the case of SiC and GaAs,
up to several tens of nanometers for SiO$_2$ and AZO,
respectively. Such variations are almost entirely due to differences
in thermal conductivities, which naturally play a major role in this
problem, with the conductivities of SiC, SiO$_2$, and AZO taken to be
$\kappa \simeq 120\,$W/m$\cdot$K, $1.4\,$W/m$\cdot$K, and
$1.2\,$W/m$\cdot$K, respectively. Note that, generally, zinc oxides
exhibit moderate values of thermal conductivities at high
temperatures, depending on their fabrication method, with the value
here taken from~\citeasnoun{loureiro2014transparent}. The open circle
in Fig.~\ref{h0} indicates the expected $\tilde{d} \sim$ hundreds of
nanometers associated with ultra-low conductivity ($\kappa\lesssim
0.05\,$W/m$\cdot$K) nanocomposite oxides that can now be
engineered~\cite{jood2011doped,chiritescu2007ultralow,loureiro2014transparent}
and which are likely to play a more prominent role in future thermal
devices~\cite{cahill2014nanoscale}. We stress that our predictions are
consistent with the lack of gradient effects observed in recent
experiments involving materials such as silicon and Au, which exhibit
low and high values of $h_0$ and $\kappa$, respectively. The case of
silica is particularly interesting, however, since it is typically
used in RHT experiments, yet the possibility of temperature gradients
has never been considered. These results along with \eqref{analytics}
can serve as a reference for future experiments, allowing estimates of
the regimes under which these effects become relevant.

While our analysis above is based on the assumption of vacuum gaps, it is straightforward to generalize Eq.~\eqref{analytics} to include the possibility of finite intervening conductivities, $\kappa_0>0$, requiring only that $h_0$ be replaced with $h_0+\kappa_0d$ in the first expression of Eq.~\eqref{analytics}. We find, however, that similar conclusions follow for small but finite $\kappa_0 \lesssim 10^{-5}\,$W/m$\cdot$K (typical of RHT experiments).

In conclusion, we have presented a study of coupled
conduction--radiation heat transfer between planar objects at short
distances. We have expressed the resulting temperature gradients and
radiative-flux modifications in terms of simple, analytical
expressions involving geometric and material parameters, showing
that in systems well within experimental reach or already considered
in experiments~\cite{ShenNanoLetters09}, both temperature gradients
and flux saturation should be observed. A similar saturation
phenomenon has been predicted to occur due to non-local
damping~\cite{Henkel06,JoulainJQSRT} and/or phonon-tunneling below the
nanometer scale~\cite{chiloyan2015natcomm} (note that at atomistic
scales where continuum electrodynamics fails, the boundary between
phonon and radiative conduction is blurred).  Our work suggests that
even at and above nano-meter gaps, and depending on material and
geometric conditions, CR interplay could instead become the dominant
mechanism limiting RHT. Furthermore, there are significant efforts
underway aimed at exploring regimes, e.g. smaller gap sizes or
materials and structures leading to larger RHT (for applications in
nanoscale cooling~\cite{cooling} and other thermal
devices~\cite{devices}), where these effects may be observed at even
at larger separations. Our ongoing work generalizing the coupled CR
formulation to arbitrary geometries reveals even larger interplay in
structured surfaces~\cite{FVC,Weiliangfuture}. Arguably, advances in
either or both directions will make such analyses necessary.

\emph{Acknowledgements} This work was supported by the National
Science Foundation under Grant no. DMR-1454836 and by the Princeton
Center for Complex Materials, a MRSEC supported by NSF Grant DMR
1420541.


\begin{thebibliography}{99}
\bibitem{howell2010thermal}J. R. Howell, M. P. Meng\"{u}\c{c}, and R. Siegel, \emph{Thermal radiation heat transfer} (CRC press, 2010).
\bibitem{basu2009review}S. Basu, Z. Zhang, and C. Fu, International Journal of Energy Research \textbf{33}, 1203 (2009).
\bibitem{volokitin2007near}A. Volokitin and B. N. Persson, Rev. Mod. Phys. \textbf{79}, 1291 (2007).
\bibitem{joulain2005review}K. Joulain, J.-P. Mulet, F. Marquier, R. Carminati, and J.-J. Greffet, Surf. Sci. Rep. \textbf{57}, 59 (2005).
\bibitem{miller2015shape}O. D. Miller, S. G. Johnson, and A. W. Rodriguez, Phys. Rev. Lett. \textbf{115}, 204302 (2015).
\bibitem{khandekar2015giant}C. Khandekar, W. Jin, O. D. Miller, A. Pick, and A. W. Rodriguez, preprint arXiv:1511.04492 (2015).
\bibitem{narayanaswamy2008thermal}A. Narayanaswamy and G. Chen, Phys. Rev. B \textbf{77}, 075125 (2008).
\bibitem{kruger2011nonequilibrium}M. Krüger, T. Emig, and M. Kardar, Phys. Rev. Lett. \textbf{106}, 210404 (2011).
\bibitem{mccauley2012modeling}A. P. McCauley, M. H. Reid, M. Krüger, and S. G. Johnson, Phys. Rev. B \textbf{85}, 165104 (2012).
\bibitem{rodriguez2013fluctuating}A. W. Rodriguez, M. H. Reid, and S. G. Johnson, Phys. Rev. B \textbf{88}, 054305 (2013).
\bibitem{holman2010heat}J. Holman, \emph{Heat transfer} (McGraw-Hill Inc, 2010).
\bibitem{wong2011monte}B. T. Wong, M. Francoeur, and M. P. Meng\"{u}\c{c}, International Journal of Heat and Mass Transfer \textbf{54}, 1825 (2011).
\bibitem{lau2016parametric}J. Z.-J. Lau, V. N.-S. Bong, and B. T. Wong, J. Quant. Spectrosc. Radiat. Transfer \textbf{171}, 39 (2016).
\bibitem{ShenNanoLetters09}S. Shen, A. Narayanaswamy, and G. Chen, Nano Letters \textbf{9}, 2909 (2009).
\bibitem{chiritescu2007ultralow}C. Chiritescu, D. G. Cahill, N. Nguyen, D. Johnson, A. Bodapati, P. Keblinski, P. Zschack, Science \textbf{315}, 351 (2007).
\bibitem{cahill2014nanoscale}D. G. Cahill, P. V. Braun, G. Chen, D. R. Clarke, S. Fan, K. E. Goodson, P. Keblinski, W. P. King, G. D. Mahan, A. Majumdar, Applied Physics Reviews \textbf{1}, 011305 (2014).
\bibitem{JoulainJQSRT}K. Joulain, J. Quant. Spectrosc. Radiat. Transfer \textbf{109}, 294 (2008).
\bibitem{chiloyan2015natcomm}V. Chiloyan, J. Garg, K. Esfarjani, and G. Chen, Nature Communications \textbf{6}, 6775 (2015).
\bibitem{baffou2010mapping}G. Baffou, C. Girard, and R. Quidant, Phys. Rev. Lett. \textbf{104}, 136805 (2010).
\bibitem{baffou2014deterministic}G. Baffou, E. B. Ureña, P. Berto, S. Monneret, R. Quidant, and H. Rigneault, Nanoscale \textbf{6}, 8984 (2014).
\bibitem{ma2014heat}H. Ma, P. Tian, J. Pello, P. M. Bendix, and L. B. Oddershede, Nano Letters \textbf{14}, 612 (2014).
\bibitem{baldwin2014thermal}C. L. Baldwin, N. W. Bigelow, and D. J. Masiello, The journal of physical chemistry letters \textbf{5}, 1347 (2014).
\bibitem{biswas2015sudden}R. Biswas and M. L. Povinelli, ACS Photonics \textbf{2}, 1681 (2015).
\bibitem{wong2014coupling}B. T. Wong, M. Francoeur, V. N.-S. Bong, and M. P. Meng\"{u}\c{c}, J. Quant. Spectrosc. Radiat. Transfer \textbf{143}, 46 (2014).
\bibitem{KittelPRL05}A. Kittel, W. M\"{u}ller-Hirsch, J. Parisi, S.-A. Biehs, D. Reddig, and M. Holthaus, Phys. Rev. Lett. \textbf{95}, 224301 (2005).
\bibitem{NarayanaswamyPRB08}A. Narayanaswamy, S. Shen, and G. Chen, Phys. Rev. B \textbf{78}, 115303 (2008).
\bibitem{HuApplPhysLett08}L. Hu, A. Narayanaswamy, X. Chen, and G. Chen, Appl. Phys. Lett. \textbf{92}, 133106 (2008).
\bibitem{RousseauNaturePhoton09}E. Rousseau, A. Siria, G. Joudran, S. Volz, F. Comin, J. Chevrier, and J.-J. Greffet, Nature Photon. \textbf{3}, 514 (2009).
\bibitem{OttensPRL11}R. S. Ottens, V. Quetschke, S. Wise, A. A. Alemi, R. Lundock, G. Mueller, D. H. Reitze, D. B. Tanner, and B. F. Whiting, Phys. Rev. Lett. \textbf{107}, 014301 (2011).
\bibitem{KralikRevSciInstrum11}T. Kralik, P. Hanzelka, V. Musilova, A. Srnka, and M. Zobac, Rev. Sci. Instrum. \textbf{82}, 055106 (2011).
\bibitem{KralikPRL12}T. Kralik, P. Hanzelka, M. Zobac, V. Musilova, T. Fort, and M. Horak, Phys. Rev. Lett. \textbf{109}, 224302 (2012).
\bibitem{vanZwolPRL12a}P. J. van Zwol, L. Ranno, and J. Chevrier, Phys. Rev. Lett. \textbf{108}, 234301 (2012).
\bibitem{vanZwolPRL12b}P. J. van Zwol, S. Thiele, C. Berger, W. A. de Heer, and J. Chevrier, Phys. Rev. Lett. \textbf{109}, 264301 (2012).
\bibitem{SongNatureNano15}B. Song \emph{et al.}, Nature Nanotechnology \textbf{10}, 253 (2015).
\bibitem{KimNature15}K. Kim \emph{et al.}, Nature \textbf{528}, 387 (2015).
\bibitem{KloppstecharXiv}K. Kloppstech \emph{et al.}, preprint arXiv:1510.06311 (2015).
\bibitem{StGelaisNatureNano16}R. St-Gelais, L. Zhu, S. Fan, and M. Lipson, Nature Nanotechnology \textbf{11}, 515 (2016).
\bibitem{ChapuisPRB08}P.-O. Chapuis, S. Volz, C. Henkel, K. Joulain, and J.-J. Greffet, Phys. Rev. B \textbf{77}, 035431 (2008).
\bibitem{MuletMTE02}J.-P. Mulet, K. Joulain, R. Carminati, and J.-J.Greffet, Microscale Thermophysical Engineering \textbf{6}, 209 (2002).
\bibitem{Henkel06}C. Henkel and K. Joulain, Appl. Phys. B 84, 61 (2006).
\bibitem{da2004conduction}R. M. S. da Gama, Applied Mathematical Modelling \textbf{28}, 795 (2004).
\bibitem{rytov1988principles}S. M. Rytov, Y. A. Kravtsov, and V. I. Tatarskii, \emph{Principles of Statistical Radiophysics} (Springer, Berlin, 1988).
\bibitem{chen2005surface}D.-Z. A. Chen, A. Narayanaswamy, and G. Chen, Phys. Rev. B \textbf{72}, 155435 (2005).
\bibitem{volz2016nanophononics}S. Volz, \emph{et al.}, The Eur. Phys. J. B \textbf{89}, 1 (2016).
\bibitem{francoeur2009solution}M. Francoeur, M. P. Meng{\"u}{\c{c}}, and. R. Vaillon, J. Quant. Spectrosc. Radiat. Transfer \textbf{110}, 2002 (2009).
\bibitem{ben2009near}P. Ben-Abdallah, K. Joulain, J. Drevillon, and G. Domingues, J. Appl. Phys. \textbf{106}, 044306 (2009).
\bibitem{MessinaPRA11}R. Messina and M. Antezza, Phys. Rev. A \textbf{84}, 042102 (2011).
\bibitem{Messina2012prl}R. Messina, M. Antezza, and P. Ben-Abdallah, Phys. Rev. Lett. \textbf{109}, 244302 (2012).
\bibitem{MessinaPRA14}R. Messina and M. Antezza, Phys. Rev. A \textbf{89}, 052104 (2014).
\bibitem{futurework}R. Messina, W. Jin, and A. W. Rodriguez, in preparation.
\bibitem{Palik98}\emph{Handbook of Optical Constants of Solids}, edited by E. Palik (Academic Press, New York, 1998).
\bibitem{loureiro2014transparent}J. Loureiro \emph{et al.}, J. Mater. Chem. A \textbf{2}, 6649 (2014).
\bibitem{jood2011doped}P. Jood, R. J. Mehta, Y. Zhang, G. Peleckis, X. Wang, R. W. Siegel, T. Borca-Tasciuc, S. X. Dou, and G. Ramanath, Nano Letters \textbf{11}, 4337 (2011).
\bibitem{cooling}B. Guha, C. Otey, C. B. Poitras, S. Fan, and M. Lipson, Nano Letters \textbf{12}, 4546 (2012).
\bibitem{devices}P. Ben-Abdallah and S.-A. Biehs, AIP Advances \textbf{5}, 053502, (2015).
\bibitem{FVC}A. G. Polimeridis, M. T. H. Reid, W. Jin, S. G. Johnson, J. K. White, and A. W. Rodriguez, Phys. Rev. B \textbf{92},134202 (2015). 
\bibitem{Weiliangfuture}W. Jin, R. Messina, and A. W. Rodriguez, in preparation.
\end{thebibliography}
\end{document}